\documentclass[12pt]{iopart}
\usepackage{graphicx}

\begin{document}

\title{A simulation of hard spheres gas for educational purposes}

\author{M Mere\v{s}$^1$ and B Tom\'a\v{s}ik$^{1,2}$}

\address{$^1$ Univerzita Mateja Bela, 97401 Bansk\'a Bystrica, Slovakia\\
$^2$ FJFI, Czech Technical University in Prague, B\v{r}ehov\'a 7, 11519~Prague,
Czech~Republic}
\ead{boris.tomasik@cern.ch, michal.meres@gmail.com}
\begin{abstract}
We describe an educational simulation of some effects within the gas of hard spheres. 
The focus of the presented simulation is on the comprehension of random 
character of the velocity of molecules in the gas and of the energy at 
fixed temperature. It allows to point out the connection between the temperature,
the energy, and the velocity of the molecules.
\end{abstract}


\section{Introduction}

Thermodynamics and statistical physics provide the key to understanding 
many physics phenomena and are therefore very important part of the secondary 
school physics curriculum. A difficulty in their thorough understanding 
together with the corresponding formulae  is in the statistical 
treatment of a large number of atoms or molecules. Quantities like 
\emph{internal energy} or \emph{mean quadratic velocity} are given by the 
\emph{averages} of those atributed to individual molecules.
Internal energy is connected with the temperature. 

The averaging is based on the underlying Maxwell-Boltzmann distribution. 
It goes beyond the secondary school curriculum to formally show, that this 
is the distribution which does not change in time under influence of 
the collisions among the molecules. Moreover, most distributions of the momenta
of molecules will converge to Maxwell-Boltzmann form as a result of
molecular collisions. 

We present here a simulation applet which explains these features of 
thermodynamics without the need of formal mathematical formulae. 
The topic of using applets for teaching physics is widely discussed 
in the literature, see e.g. \cite{christian} and \cite{wieman} and references therein.
There are many 
simulations and applets, some also available on the web, e.g.~\cite{phet,natsim,taiwan}.
The teacher can choose among them if he or she wants to support the explanation
in the classroom or refer the pupils to some supporting material. It is a matter 
of appropriatness to the particular topic which is being explained. In our 
simulations we have put emphasis on i) explaining the correspondence between 
gas energy and mean quadratic velocity calculated from the temperature and those
really measured---if this was possible on individual molecules; and ii) explain 
the random nature of the velocities of the molecules. 

Our simulations are freely available on the web \cite{ourweb}


\section{Ideal gas relations}

The simulation shows \emph{two-dimensional} gas of 500 single-atom molecules. 
Collisions of atoms are treated as elastic collisions of circles; this determines
the angle of scattering and the outgoing momenta. Thus, strictly speaking, the gas is 
not ideal but rather described by the van der Waals equation of state. It is dilute, 
however, so that ideal gas relations can be applied. Collisions with the walls of the 
container are elastic, except for one wall which can be ``heated''. 

We chose to perform the simulation in two dimensions since in such a case we can 
show all molecules straightforwardly and do not need to do any projections to a plane, 
which might be confusing. On the other hand, we have to keep in mind that this 
changes some relations. The (mean) internal energy of the gas is 
\begin{equation}
E = \frac{2}{2} N\, k_B\, T\, ,
\end{equation}
where $N$ is the number of atoms, $T$ is the temperature and $k_B$ the Boltzmann constant.
Intentionally we display the factor $2/2$---instead of $3/2$---indicating that we are in 
two dimensions. Correspondingly, the relation for mean quadratic velocity reads
\begin{equation}
\label{meanvel}
v = \sqrt{\frac{2\, k_B\, T}{m}}\,  ,
\end{equation}
where $m$ is the mass of the atom and the factor 2 is due to two dimensions. Velocities are  
distributed according to the Maxwell distribution, which in two dimensions takes the form
\begin{equation}
\label{md}
{\cal P}(v_x,v_y) = \frac{m}{2\pi\, k_B\, T} \exp\left ( - \frac{m(v_x^2 + v_y^2)}{2\, k_B\, T}  \right ) \, .
\end{equation}


\section{The simulation}

\begin{figure}[t]
\begin{center}
\includegraphics[width=0.7\textwidth]{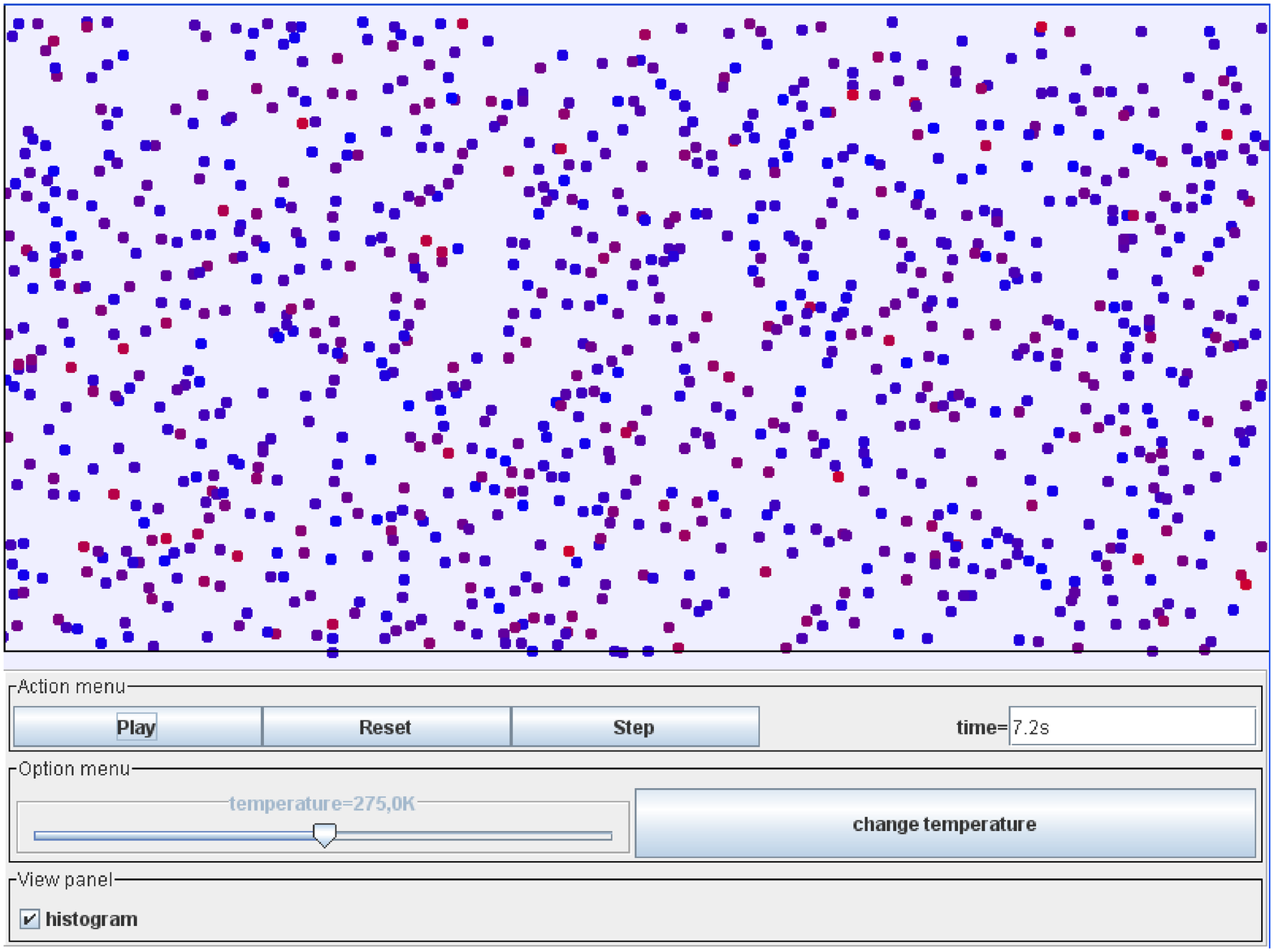}
\end{center}
\caption{%
A typical simulation at high temperature; here the temperature is set to 500~K. At lower temperature 
there are less red (fast) atoms. 
}
\label{f:gas}
\end{figure}

The simulation always shows a rectangular container with the gas. 
The color of the atom is chosen from a spectrum 
between blue (slow atoms) and red (fast atoms). In Figure~\ref{f:gas} 
we show how the simulation looks like. 
Three of the sides always bounce off the atoms elastically. 

One side can be made into a heater/cooler. In that case, its temperature 
can be set to some value.  When the atoms bounce off this side, their velocity is determined randomly 
according to Maxwell distribution (eq.~(\ref{md})) with the temperature parameter set to that of the 
heater/cooler and direction inwards the containter. 

In a different setting, one of the sides is replaced by a piston which can be moved eigter by pressing 
arrow buttons in the simulation window or by clicking a draging.

Additional windows can be displayed. One of them contains 
the histogram of the velocity distribution which is 
updated in real time. Another one shows the graph of the time dependence of the gas energy 
compared with the value which 
the energy is expected to have according to the temperature of the heated 
wall. In the last auxilliary window the values of the actual energy, expected energy and mean 
quadratic velocity are displayed. The temperature ranges between 50 and 500~K and the typical velocities 
are of the order of hundreds of m.s$^{-1}$. With 500 atoms the typical energy is of the order of few 
electronvolts. 


\section{The presented effects}

\subsection{Velocity distribution}

\begin{figure}[t]
\begin{center}
\includegraphics[width=0.325\textwidth]{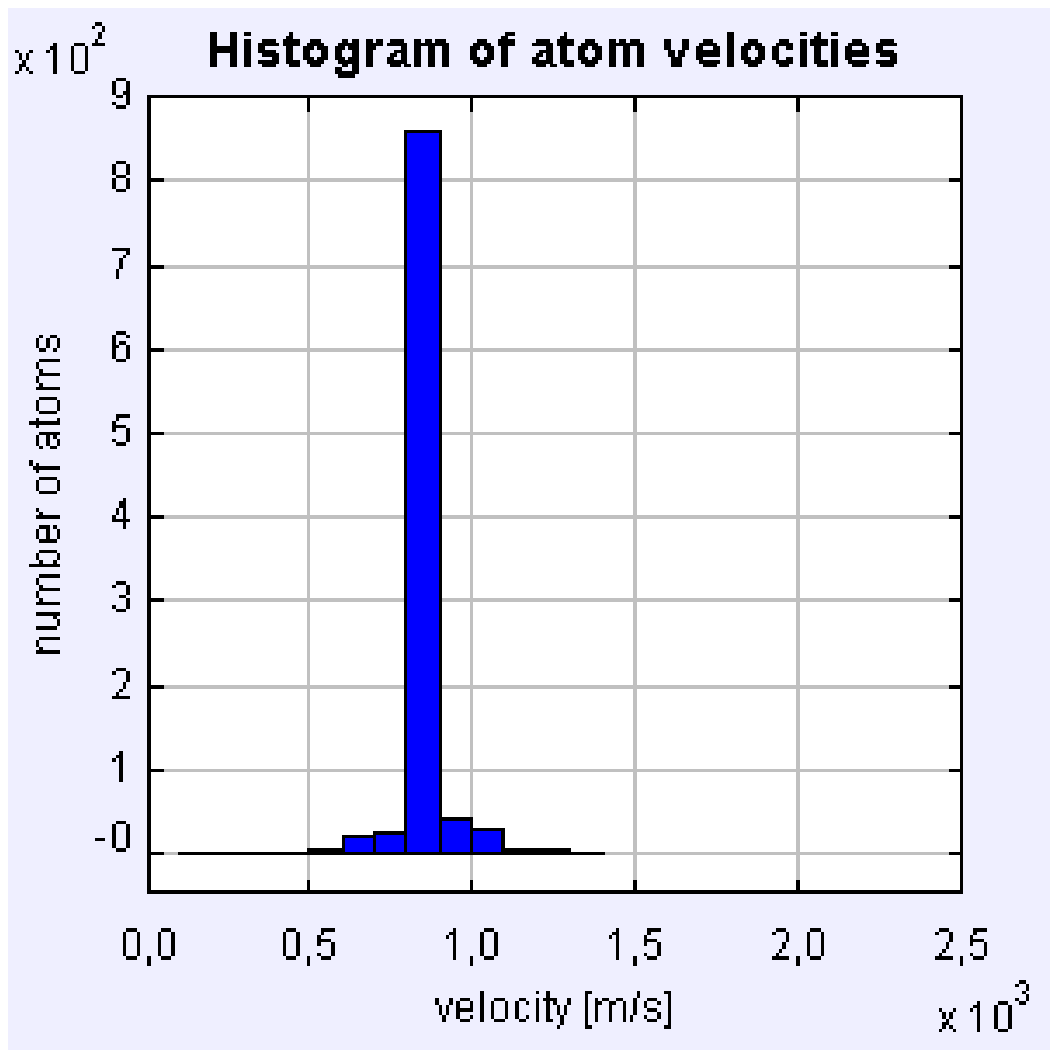}
\includegraphics[width=0.325\textwidth]{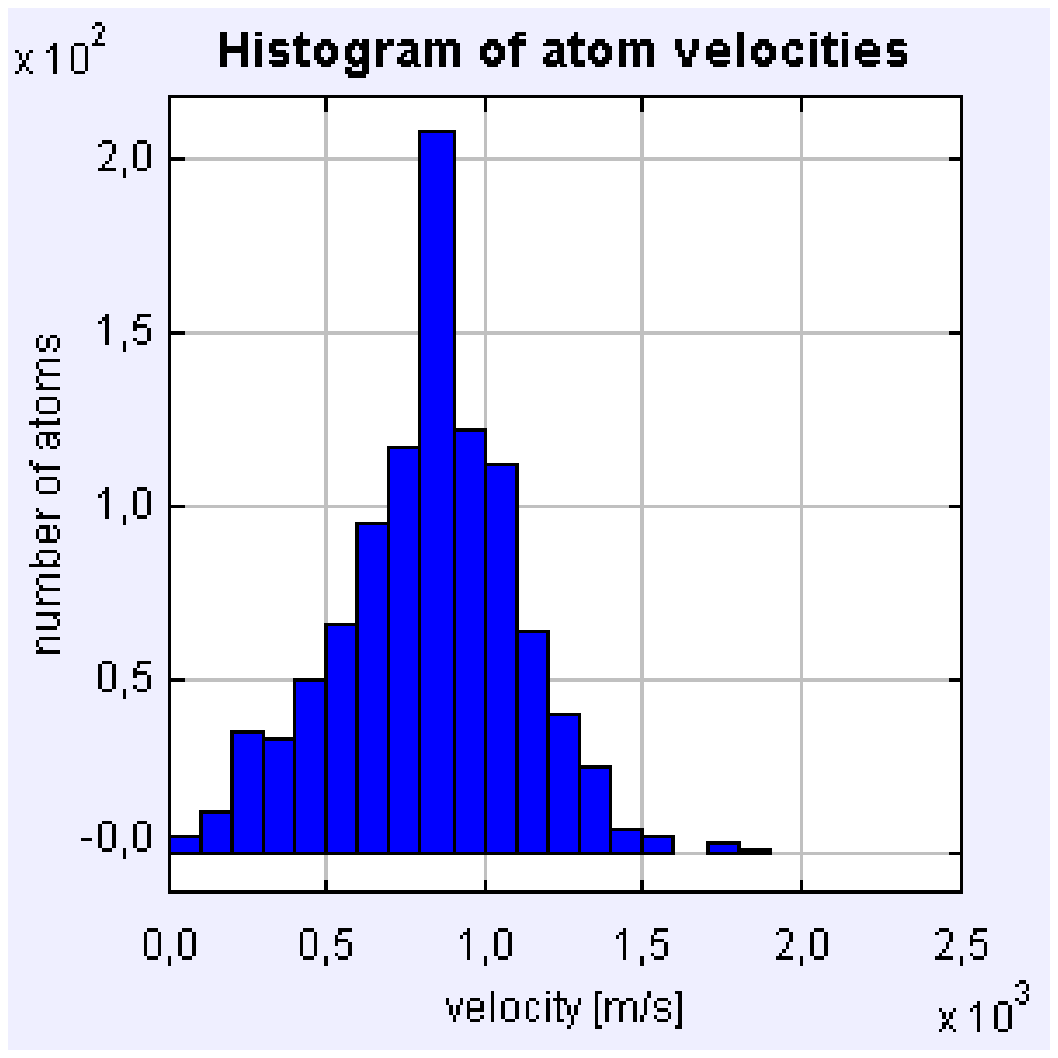}
\includegraphics[width=0.325\textwidth]{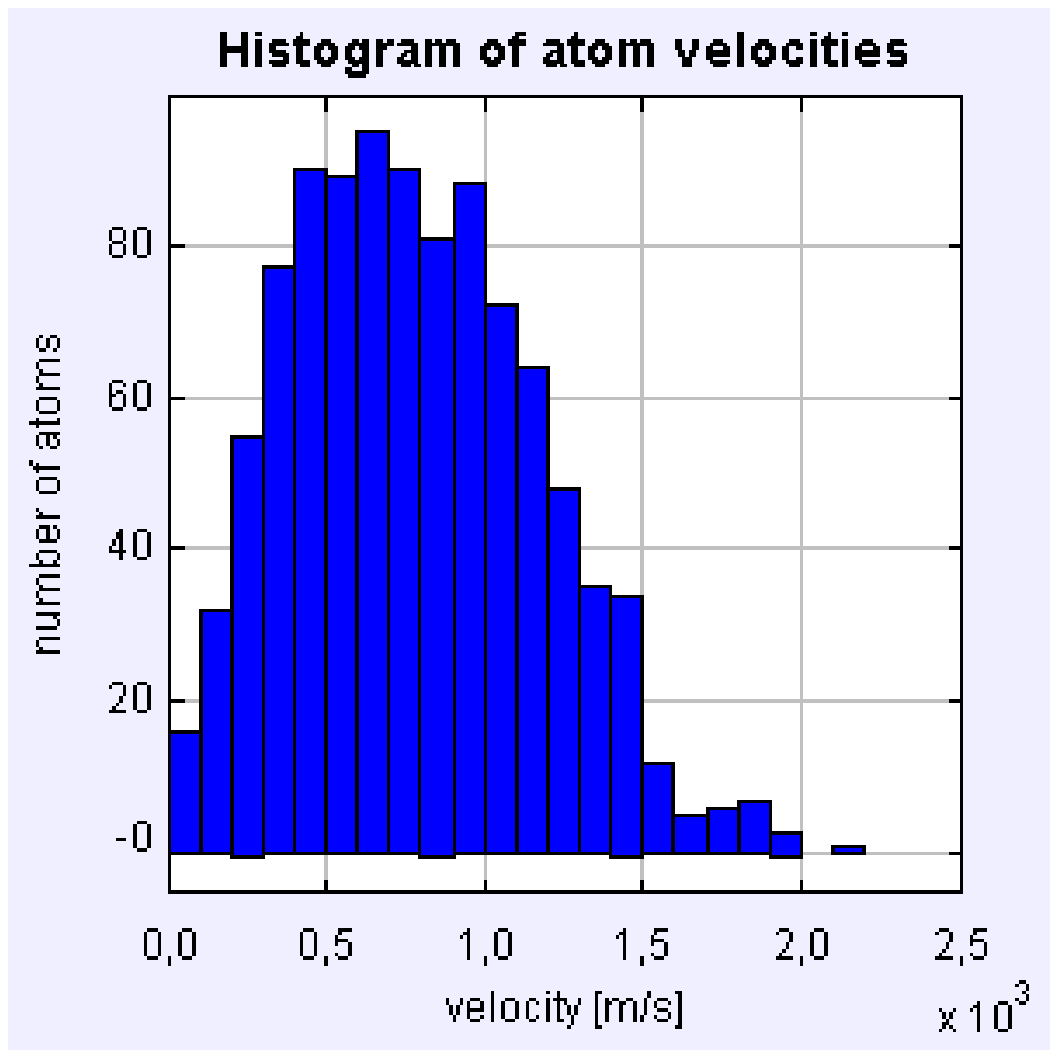}
\end{center}
\caption{%
The influence of collisions of the distribution of velocities. Left is the initial 
distribution at time 0, in the middle is the histogram after 0.5~s, on the right 
the distribution after 2~s. After 2~s the histogram fluctuates but does not change 
its gross shape, if the temperature is kept constant. 
}
\label{f:th}
\end{figure}

The simulation can be initiated in such a way that each atom has exactly the velocity $v$
given by eq.~(\ref{meanvel}), with random direction. During initialisation one time step 
is effectively performed and some collisions can occur, thus some velocities may slightly 
change. Nevertheless, the initial velocity histogram shows just one very narrow peak 
(Figure~\ref{f:th} left). 
The simulation can then be started and the user can observe how the velocity distribution 
quickly broadens and becomes maxwellian as a result of the collisions, see Fig.~\ref{f:th}. It is interesting 
to run the simulation step by step since the process is very quick.


\subsection{Heating and cooling}

\begin{figure}[t]
\begin{center}
\includegraphics[width=0.7\textwidth]{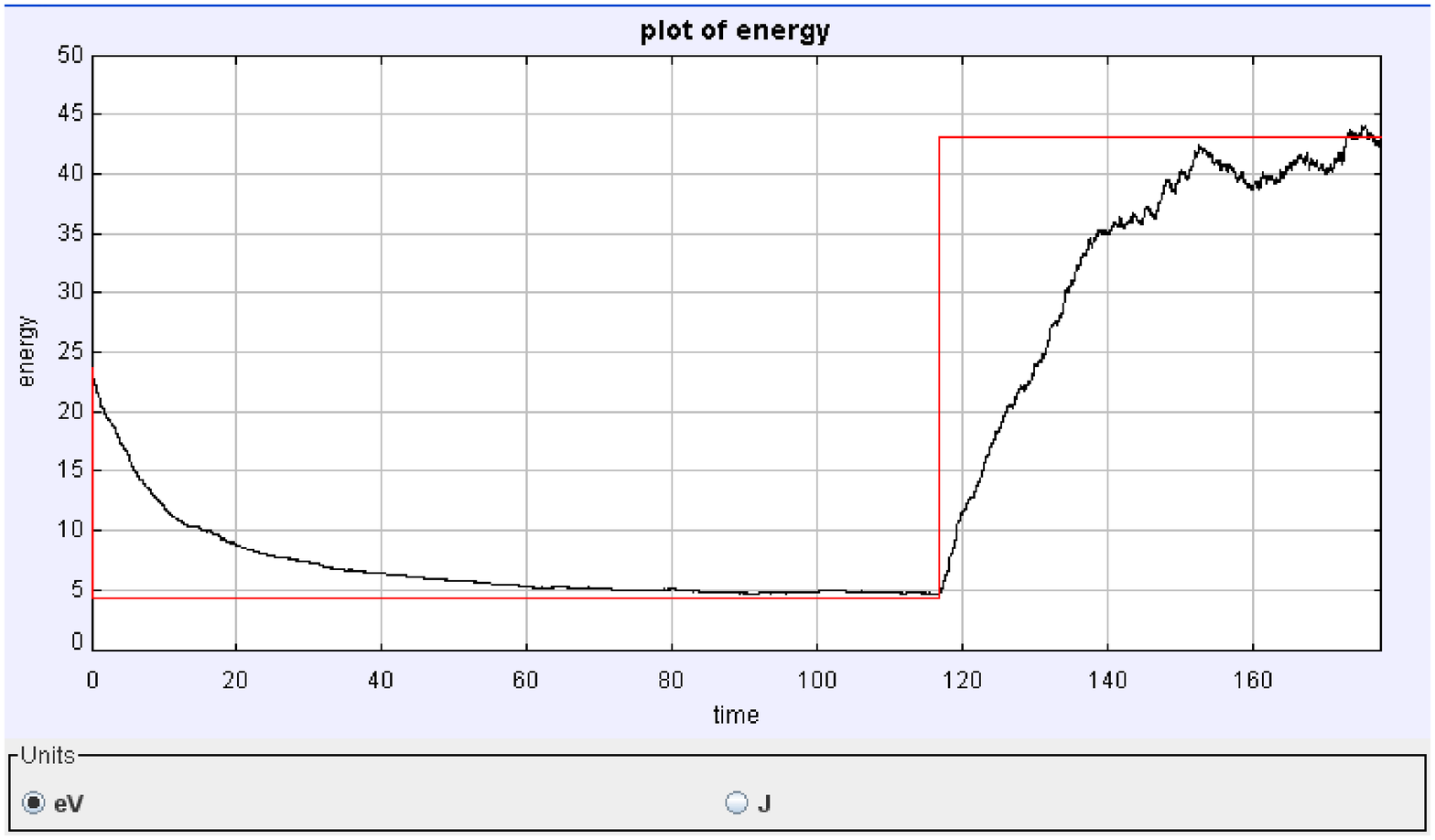}
\end{center}
\caption{%
The red straight line shows the expected energy of the gas corresponding to the temperature
of the heating wall. The black fluctuating line is the actual energy as a function of the time.
The original temperature of the gas was 250~K. In the beginning the temperature of the heating/cooling
wall was set to 50~K. After 100~s it was put to the value of 500~K. 
}
\label{f:enrg}
\end{figure}

In this mode of the simulation, one of the walls of the container can be set to a temperature 
different from that of the gas. Then is is interesting to observe the plot of the time dependence 
of the gas energy, which we show in Figure~\ref{f:enrg}. 
It shows the expected value corresponding to the temperature of the heating
or cooling wall, and the actual energy of the gas obtained by summing up kinetic energies of 
all atoms. On average, the actual energy converges to the expected one, and thermal fluctuations
are clearly visible. 

\begin{figure}[t]
\begin{center}
\includegraphics[width=0.33\textwidth]{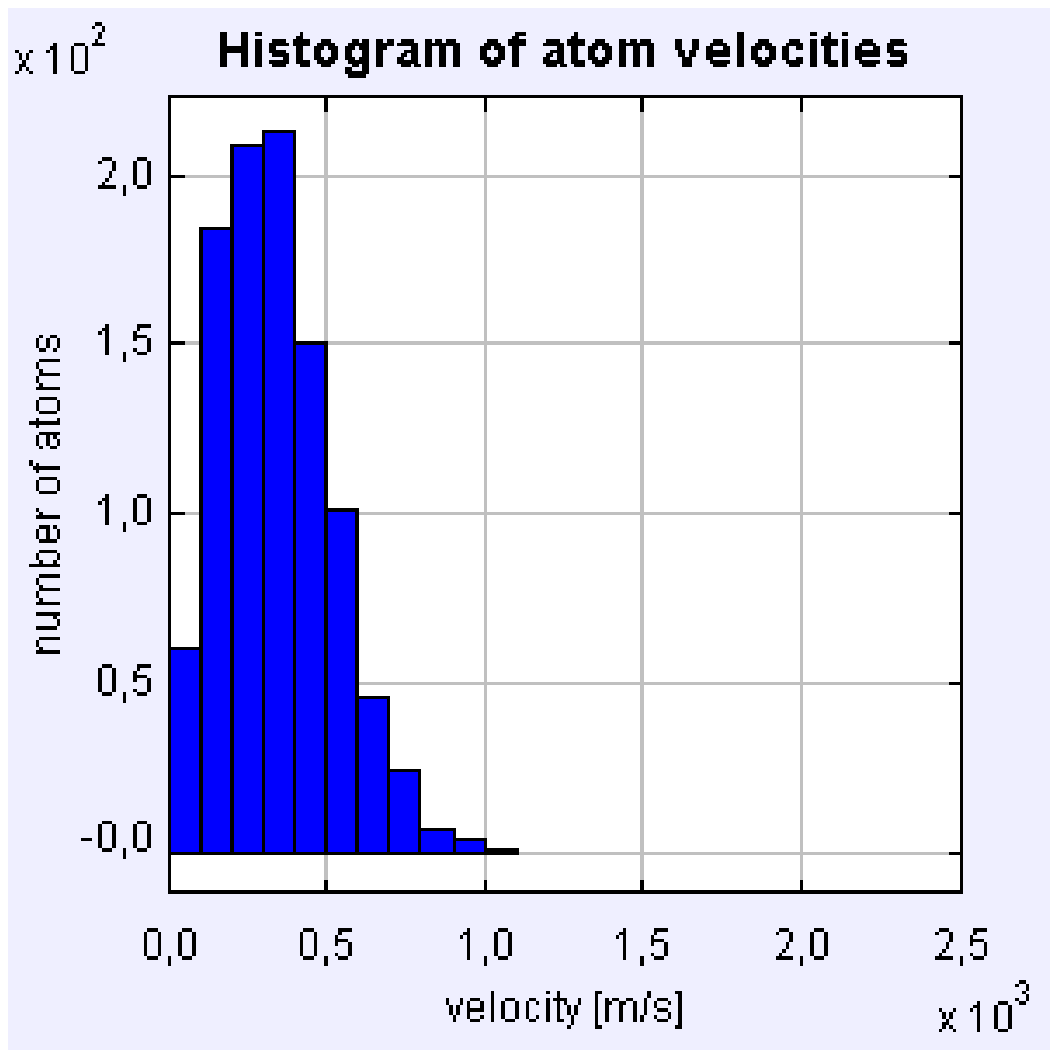}
\includegraphics[width=0.33\textwidth]{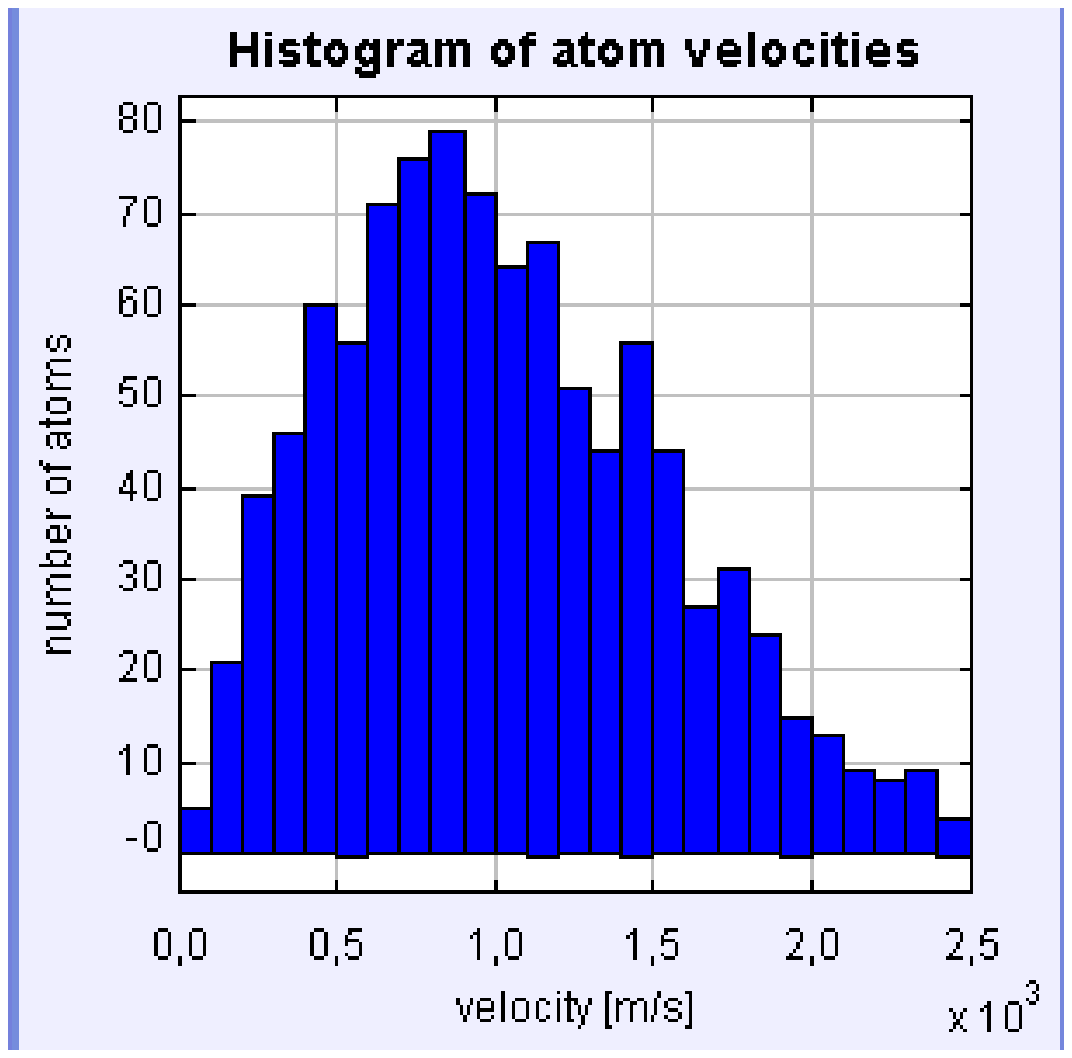}
\end{center}
\caption{%
The histograms of velocity distribution at the temperature of 50~K (left) and 500~K (right).
}
\label{f:hth}
\end{figure}

It is also interesting to observe the velocity histogram (Figure~\ref{f:hth}). 
Here one can realise
the connection between the temperature, the energy, and the distribution of velocities. It can be
seen how the distribution broadens and the peak of the histogram (the most probable velocity)
moves to higher  values with increasing temperature, and vice versa, lower temperature 
leads to narrower histogram with a peak at lower velocity.


\subsection{Pressing}

One can also demonstrate the effect of pressing and releasing of the gas. One of the 
sides of the container is then turned into a piston. The piston should be moved slowly---slower
than the typical velocity of the atoms. Otherwise the presentation is somewhat 
unrealistic. One can see that when the piston is pressing the volume, atoms are bounced 
off it with higher velocity than they had before and this increases the energy of the 
gas. This can be observed in the plot of energy as a function of time and on the velocity histogram. 
Increasing the volume with the help of the piston leads to the inverse effect.


\subsection{Brownian motion}

\begin{figure}[t]
\begin{center}
\includegraphics[width=0.8\textwidth]{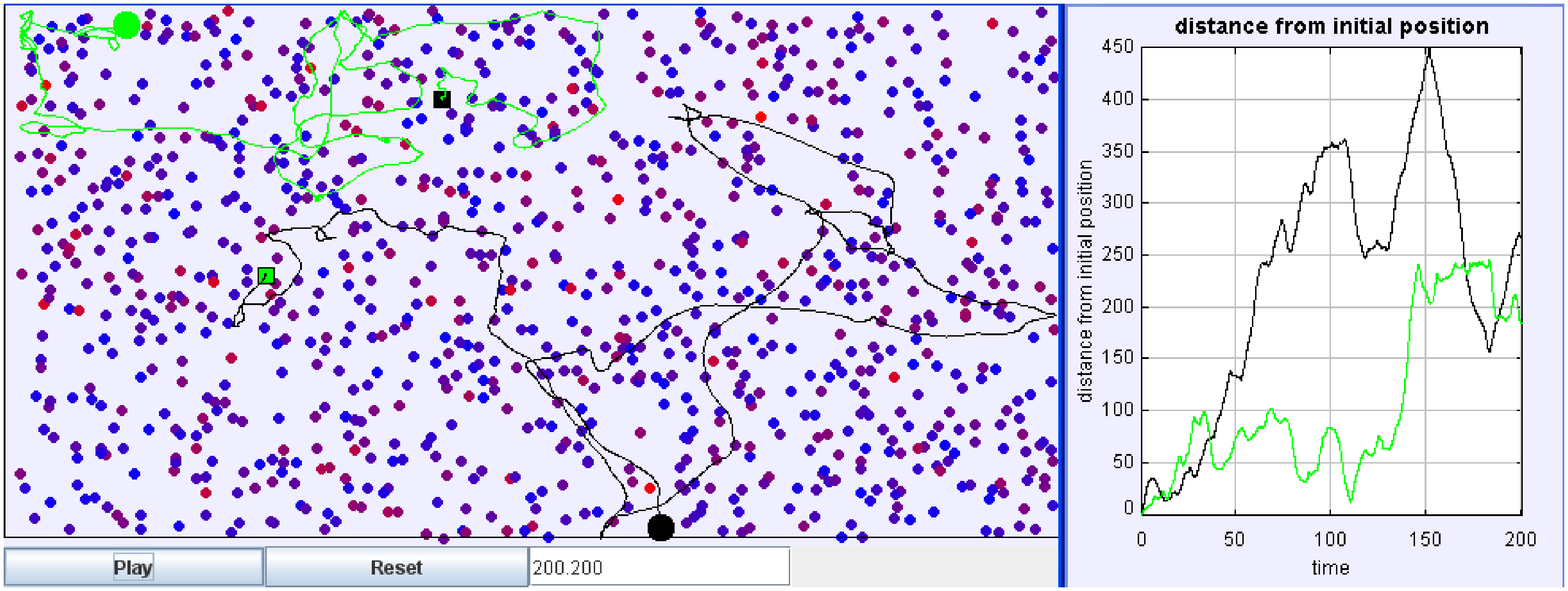}
\end{center}
\caption{%
The simulation of Brownian motion. Paths of the two heavy particles are shown. In the separate 
window, their distance from the original position as a function of time is presented. 
}
\label{f:brown}
\end{figure}

We have also prepared a simulation of the Brownian motion. Two larger circles with  masses hundred 
times larger than the atoms are immersed into the gas which is initiated at the temperature of 250~K. 
As a result of the collisions they begin to move. The trajectory they have passed is shown in the 
simulation (Figure~\ref{f:brown}). 
One can also monitor their distance from the initial position 
as a function of time  in a dedicated window.


\section{Conclusions}

We believe that our simulations can serve as a valuable supporting tool for explaining
some features of the ideal gas. We plan to enlarge the collection of the addressed effects
in the future.


\end{document}